\documentclass[%
reprint,
superscriptaddress,
amsmath,amssymb,amsthm,mathrsfs,
aps,
prl,
]{revtex4-1}

\usepackage{graphicx}% Include figure files
\usepackage{dcolumn}% Align table columns on decimal point
\usepackage{bm}% bold math
\usepackage[usenames]{color}
\usepackage[ %showframe,%Uncomment any one of the following lines to test
margin=0.75in,
]{geometry}
\usepackage[normalem]{ulem}

\graphicspath{{./figs/}}
\def\TN{$T_\mathrm{N}$}

\def\muB{$\mu_{\mathrm{B}}$}
\def\deltampdf{3D-$\Delta$mPDF}
\def\xic{$\xi_c$}
\def\xiab{$\xi_{ab}$}

\begin{document}
	
\title{
	Real-space visualization of short-range antiferromagnetic correlations in a magnetically enhanced thermoelectric
}

	\author{Raju Baral}
	\affiliation{ %
		Department of Physics and Astronomy, Brigham Young University, Provo, Utah 84602, USA.
	} %
	
	\author{Jacob Christensen}
	\affiliation{ %
		Department of Physics and Astronomy, Brigham Young University, Provo, Utah 84602, USA.
	} %
	
	\author{Parker Hamilton}
	\affiliation{ %
		Department of Physics and Astronomy, Brigham Young University, Provo, Utah 84602, USA.
	} %

	\author{Feng Ye}
	\affiliation{ %
		Neutron Scattering Division, Oak Ridge National Laboratory, Oak Ridge, Tennessee 37831, USA.
	} %
	
	\author{Karine Chesnel}
	\affiliation{ %
		Department of Physics and Astronomy, Brigham Young University, Provo, Utah 84602, USA.
	} %

	\author{Taylor D. Sparks}
	\affiliation{ %
		Department of Materials Science and Engineering, University of Utah, Salt Lake City, Utah 84112, USA.
	} %

	\author{Rosa Ward}
	\affiliation{ %
		Department of Materials Science and Engineering, University of Utah, Salt Lake City, Utah 84112, USA.
	} %

	\author{Jiaqiang Yan}
	\affiliation{ %
		Materials Science \& Technology Division, Oak Ridge National Laboratory, Oak Ridge, Tennessee 37831, USA.
	} %
	
	\author{Michael A. McGuire}
	\affiliation{ %
		Materials Science \& Technology Division, Oak Ridge National Laboratory, Oak Ridge, Tennessee 37831, USA.
	} %
	
	\author{Michael E. Manley}
	\affiliation{ %
		Materials Science \& Technology Division, Oak Ridge National Laboratory, Oak Ridge, Tennessee 37831, USA.
	} %
	
	\author{Julie B. Staunton}
	\affiliation{ %
	Department of Physics, University of Warwick, Coventry CV4 7AL, United Kingdom.
	}%
	
	\author{Rapha\"el P. Hermann}
	\email{hermannrp@ornl.gov}
	\affiliation{ %
		Materials Science \& Technology Division, Oak Ridge National Laboratory, Oak Ridge, Tennessee 37831, USA.
	} %

	\author{Benjamin A. Frandsen}
	\email{benfrandsen@byu.edu; lead contact}
	\affiliation{ %
		Department of Physics and Astronomy, Brigham Young University, Provo, Utah 84602, USA.
	} %

\begin{abstract}
Short-range magnetic correlations can significantly increase the thermopower of magnetic semiconductors, representing a noteworthy development in the decades-long effort to develop high-performance thermoelectric materials. Here, we reveal the nature of the thermopower-enhancing magnetic correlations in the antiferromagnetic semiconductor MnTe. Using magnetic pair distribution function analysis of neutron scattering data, we obtain a detailed, real-space view of robust, nanometer-scale, antiferromagnetic correlations that persist into the paramagnetic phase above the N\'eel temperature \TN\ = 307 K. The magnetic correlation length in the paramagnetic state is significantly longer along the crystallographic \textit{c} axis than within the \textit{ab} plane, pointing to anisotropic magnetic interactions. \textit{Ab initio} calculations of the spin-spin correlations using density functional theory in the disordered local moment approach reproduce this result with quantitative accuracy. These findings constitute the first real-space picture of short-range spin correlations in a magnetically enhanced thermoelectric and inform future efforts to optimize thermoelectric performance by magnetic means.\footnote{This manuscript has been authored by UT-Battelle, LLC under Contract No. DE-AC05-00OR22725 with the U.S. Department of Energy. The United States Government retains and the publisher, by accepting the article for publication, acknowledges that the United States Government retains a non-exclusive, paid-up, irrevocable, world-wide license to publish or reproduce the published form of this manuscript, or allow others to do so, for United States Government purposes. The Department of Energy will provide public access to these results of federally sponsored research in accordance with the DOE Public Access Plan (http://energy.gov/downloads/doe-public-access-plan).} 
\end{abstract}
	
\maketitle

% Matter journal MAP scale 2-Benchmark: Basic principles and material properties/structures quantified, observed, explained for the first time for a particulary system, but not intrinsically novel.

\section{Introduction}	
Despite the tremendous promise of thermoelectric devices for environmentally friendly energy applications ranging from waste heat recovery to solid state refrigeration~\cite{gault;cm13,twaha;rser16,mao;nm21}, relatively few material systems have been identified that possess excellent thermoelectric properties and are also economically and environmentally viable for widespread use. The material parameters that determine thermoelectric performance are encapsulated by the dimensionless figure of merit $zT = \sigma S^2 T/\kappa$, where $\sigma$ is the electrical conductivity, $S$ is the Seebeck coefficient (i.e., thermopower), $T$ is temperature, and $\kappa$ is the thermal conductivity. $S, \sigma$, and $\kappa$ are typically interdependent in such a way that makes it difficult to optimize all of them simultaneously; hence, the relative scarcity of high-performance thermoelectrics. %Material design principles such as the ``electron crystal, phonon glass'' concept have led to improvements in $zT$ in many materials, but additional routes toward superior thermoelectric performance may be required to unlock the full potential of thermoelectric devices.

One route toward higher $zT$ values is to exploit the magnetic degree of freedom in metals and semiconductors with unpaired electrons. An example of this is magnon drag, wherein a flux of magnons (i.e. spin waves, or thermal excitations of the long-range magnetic ordering pattern) drags charge carriers through the lattice via exchange of linear momentum, thereby enhancing the thermopower~\cite{baily;pr62,blatt;prl67,polas;jmchemc20}. The requirement to have long-range magnetic order that supports well-defined magnons would appear to limit the applicability of magnon drag, particularly since magnetic order is typically weakened at high temperatures, where thermoelectric devices are most likely to be useful. However, it was recently shown that the thermopower can be enhanced even by \textit{short-range} magnetic correlations, rather than true long-range order, through ``paramagnon drag''~\cite{zheng;sadv19,polas;jmchemc20}. Analogous to magnons, paramagnons are thermal excitations in a partly correlated paramagnetic state, i.e. a state where the magnetic moments maintain short-range correlations even in the absence of long-range magnetic order. If the local magnetic correlations have a sufficiently long length and lifetime, paramagnons will look like magnons on the time- and length-scales of their interactions with itinerant charge carriers, allowing a paramagnon flux to enhance the thermopower in like fashion as magnon drag. Thus, magnetic enhancement of $zT$ through drag effects is much more widely applicable than initially supposed, encompassing not only the relatively few materials with a well-ordered magnetic state at very high temperatures, but also the class of materials with lower-temperature magnetic transitions that nevertheless retain significant short-range magnetic correlations to elevated temperatures, e.g., as afforded by large exchange interactions.

Enhancing the thermopower through paramagnon drag was first demonstrated in the high-performance thermoelectric candidate MnTe~\cite{zheng;sadv19}, an antiferromagnetic (AF) semiconductor with the hexagonal NiAs structure type. The Mn$^{2+}$ (\textit{S} = 5/2) magnetic moments in MnTe order antiferromagnetically below \TN\ $\sim$ 307~K~\cite{kunit;jdp64}, such that the spins align ferromagnetically within hexagonal sheets and antiferromagnetically between sheets~\cite{szusz;prb06}. When lightly doped with Li, Na, or Ag, MnTe displays outstanding thermoelectric properties with a thermoelectric figure of merit \textit{zT} reaching 1 around 800 - 900~K~\cite{xu;jmchema17,ren;jmchemc17,dong;jmchemc18,zheng;sadv19}, which represents a 300\% increase in \textit{zT} over the expected value if spin effects were absent~\cite{polas;crps21}. %From a more fundamental perspective, MnTe has attracted interest as a ``crossroads'' material~\cite{allen;ssc77,youn;pssb04} because it shares characteristics of both strongly correlated Mott or charge-transfer insulators (e.g., narrow \textit{d} bands) and also weakly correlated band insulators (e.g., significant overlap of \textit{p} and \textit{d} orbitals), giving rise to rich and varied electronic and magnetic properties.

The discovery of paramagnon drag in MnTe demonstrated the viability of exploiting the spin degree of freedom to boost thermoelectric performance in novel ways~\cite{polas;crps21}. However, significant work remains to be done to benchmark the paramagnon drag effect and establish a comprehensive understanding of its role in MnTe and other materials. Specifically, a real-space picture of the short-range AF correlations is lacking, as is a theoretical framework based on first principles that can explain the observed short-range magnetism and identify other candidate systems with promising paramagnon properties. Addressing these knowledge gaps will greatly enhance future efforts to harness paramagnon effects for thermoelectricity and other functionalities.

Here, we combine cutting-edge neutron total scattering techniques with sophisticated \textit{ab initio} calculations to achieve a comprehensive picture of the short-range AF correlations underlying the paramagnon drag effect in MnTe. Using both three-dimensional (3D) and one-dimensional (1D) magnetic pair distribution function (PDF)~\cite{egami;b;utbp12,frand;aca14,frand;aca15,roth;iucrj18} analysis of neutron total scattering data, we present the first real-space visualization of nanometer-scale spin correlations in the paramagnetic state of MnTe and reveal a strongly anisotropic magnetic correlation length. These observations are quantitatively reproduced using density functional theory (DFT) with the self-interaction-corrected (SIC) local spin density approximation in the disordered local moment (DLM) approach (DLM-DFT-SIC)~\cite{gyorf;jppfm85,hughe;njp08}. These results deepen our understanding of paramagnon phenomena in MnTe and related materials and highlight the value of combining magnetic PDF (mPDF) and DLM-DFT-SIC theory to establish the nature of short-range magnetic correlations in exquisite detail. The success of this approach for MnTe provides a valuable benchmark for future applications to other materials with magnetically enhanced functionalities.

\section{Results}	
\subsection{Single-crystal neutron diffraction: 3D-$\Delta$mPDF analysis}
\begin{figure}
	\includegraphics[width=75mm]{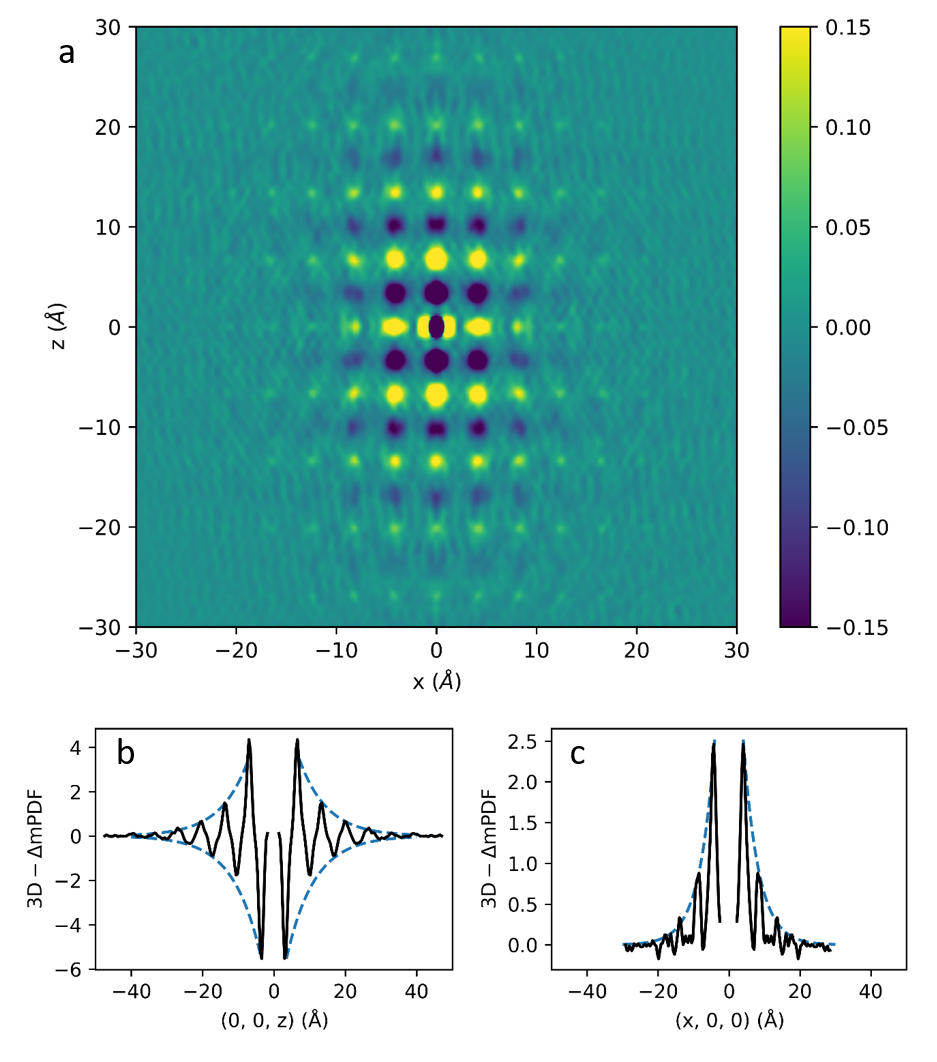}
	\caption{\label{fig:3DmPDF} (a) \deltampdf\ pattern showing short-range, anisotropic AF correlations in the \textit{xz} plane of MnTe at $T \sim 340$~K. (b) Cut through the \deltampdf\ data along the $z$ direction with $x=y=0$. Dashed curves illustrate the best-fit exponential envelope with a correlation length of 7.7(4)~\AA. (c) Same as (b), but with the cut taken along the $x$ direction. The best-fit correlation length is 4.3(2)~\AA\ along this direction. 
	}		
\end{figure}

We begin by examining the single-crystal neutron diffraction and three-dimensional difference mPDF (\deltampdf~\cite{roth;iucrj18}) data to gain insight into the short-range AF correlations above \TN. In Fig.~\ref{fig:3DmPDF}(a),
%%%%%%%%%%%%%%%%%
% Begin Figure
%%%%%%%%%%%%%%%%%
%%%%%%%%%%%%
% End Figure
%%%%%%%%%%%%
we display the \deltampdf\ at $T \sim 340$~K in the \textit{xz} plane, such that the \textit{a} (\textit{c}) crystallographic direction is parallel to the horizontal (vertical) axis [see the Supplementary Information (SI) for a view of the crystal structure]. This figure provides a direct view of the short-range AF correlations in the paramagnetic state of MnTe in real space. Moving vertically up and down the pattern, we observe alternating dark and bright regions indicative of antiferromagnetic correlations along \textit{c}, while the horizontal direction exhibits rows with uniform color, indicative of ferromagnetic correlations in the \textit{ab} plane. Thus, the short-range magnetic correlations are qualitatively similar to those in the long-range ordered state. Interestingly, Fig.~\ref{fig:3DmPDF}(a) reveals a striking anisotropy in the correlation length, with correlations along \textit{z} clearly visible to about $\pm$30~$\mathrm{\AA}$, while those along \textit{x} have a much shorter spatial extent. This can be quantified by taking cuts of the displayed data along the \textit{x} = 0 and \textit{z} = 0 lines and fitting an exponential envelope to the correlation profile, as shown in Fig.~\ref{fig:3DmPDF}(b, c). This procedure yields correlation lengths of 7.7(4)~\AA\ and 4.3(2)~\AA, respectively. We note that this \deltampdf\ pattern was collected without any energy discrimination on the instrument and therefore probes the instantaneous magnetic correlations. When just the neutrons scattered within the elastic bandwidth of $~\sim2$~meV are selected (hence providing sensitivity to the correlations with a lifetime longer than $\sim$2~ps), the out-of-plane and in-plane correlation lengths are 12.4(9)~\AA\ and 5.7(2)~\AA, respectively. Additional details regarding the \deltampdf\ analysis are given in the SI, including simulated \deltampdf\ patterns from models of the short-range magnetic order.

\subsection{Neutron powder diffraction: One-dimensional atomic and magnetic PDF analysis}
\begin{figure}
	\includegraphics[width=80mm]{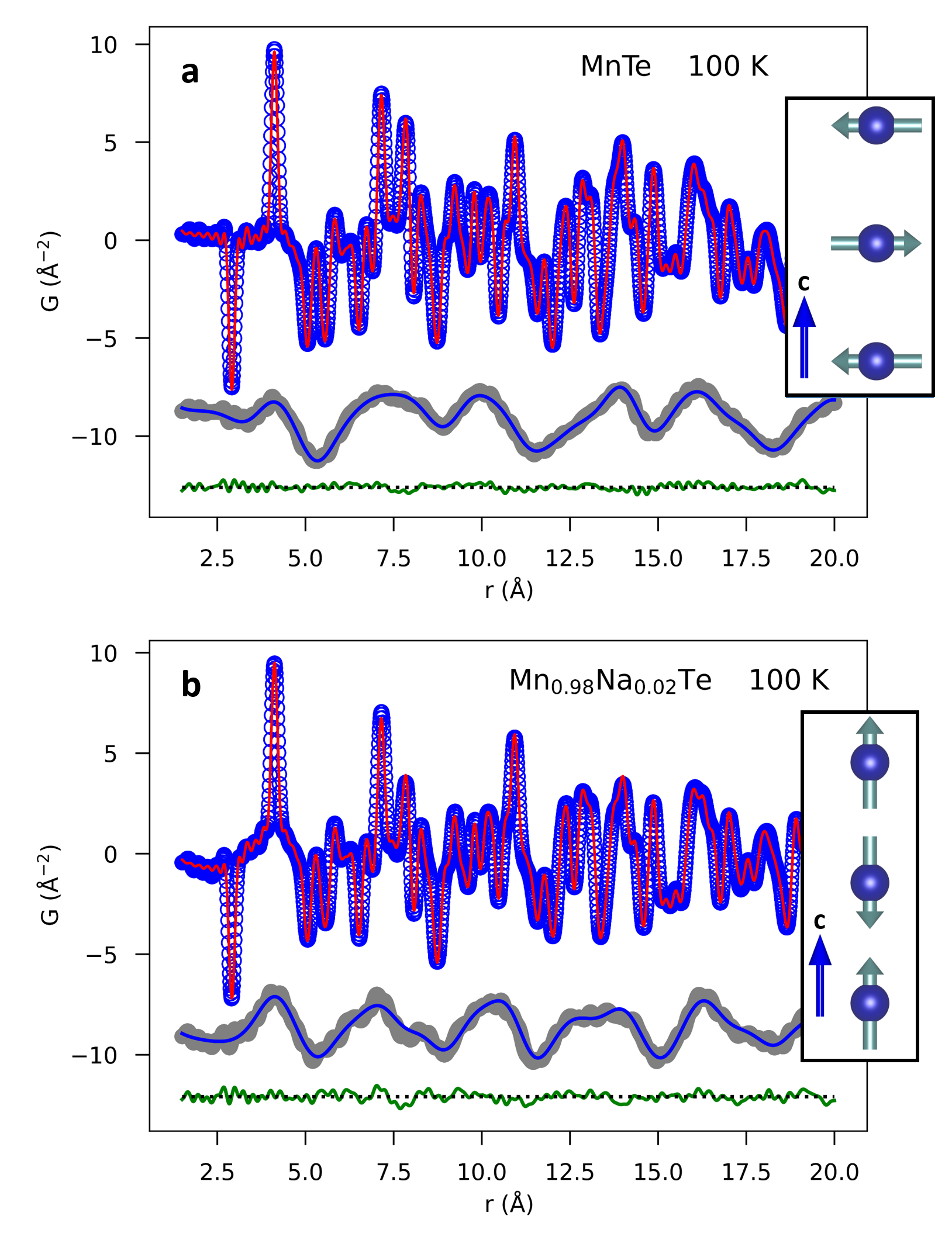}
	\caption{\label{fig:mPDFs} Combined atomic and magnetic PDF analysis for pure MnTe (a) and Na-doped MnTe (b) at 100~K, showing a strong magnetic signal and a reorientation of the sublattice magnetization from in the plane in pure MnTe to out of the plane in doped MnTe. The mPDF data (gray curve; assumed to be the difference between the observed total PDF and the calculated atomic PDF) and the calculated mPDF (solid blue curve) are offset vertically below the total PDF data and fit for clarity. Insets: Corresponding Mn spin orientations along the \textit{c} axis.
	}		
\end{figure}

We further performed one-dimensional atomic and magnetic PDF analysis of neutron total scattering data collected from powder samples of MnTe and Na-doped MnTe on a dense temperature grid across \TN. Fig.~\ref{fig:mPDFs}(a) shows the atomic and magnetic PDF fits together at 100~K for pure MnTe.
%%%%%%%%%%%%%%%%%
% Begin Figure
%%%%%%%%%%%%%%%%%
%%%%%%%%%%%%
% End Figure
%%%%%%%%%%%%
The upper set of curves represent the total PDF data (blue symbols) and total calculated PDF (red curve), which includes the best-fit atomic and magnetic PDF signals together. We used the published atomic and magnetic structures~\cite{szusz;prb06} as the starting point for the refinement (see the SI for further details). The fit is excellent ($R_w = 0.0576$), as seen by the small and featureless fit residual shown as the green curve at the bottom of the panel. The best fit occurs when the spins are aligned within the \textit{ab} plane. Due to the hexagonal symmetry, no preferred direction within the \textit{ab} plane could be detected, although Kriegner \textit{et al.}~\cite{krieg;nc16} have established that the moments lie along nearest-neighbor Mn-Mn directions in the plane. Fig.~\ref{fig:mPDFs}(b) shows the same results for the Na-doped sample. We note that the best-fit spin orientation for the Na-doped sample is along the \textit{c} axis, which accounts for the different shape of the mPDF signal compared to the pure sample and is confirmed by inspection of the magnetic Bragg peak intensities in the diffraction data, as shown in the SI. The different spin orientations for pure and doped MnTe at 100~K are illustrated in the insets of Fig.~\ref{fig:mPDFs}. These findings agree with neutron diffraction data on Li-doped MnTe~\cite{mosel;prm21} and point to a delicate balance between competing ground states in MnTe.

\begin{figure*}
	\includegraphics[width=155mm]{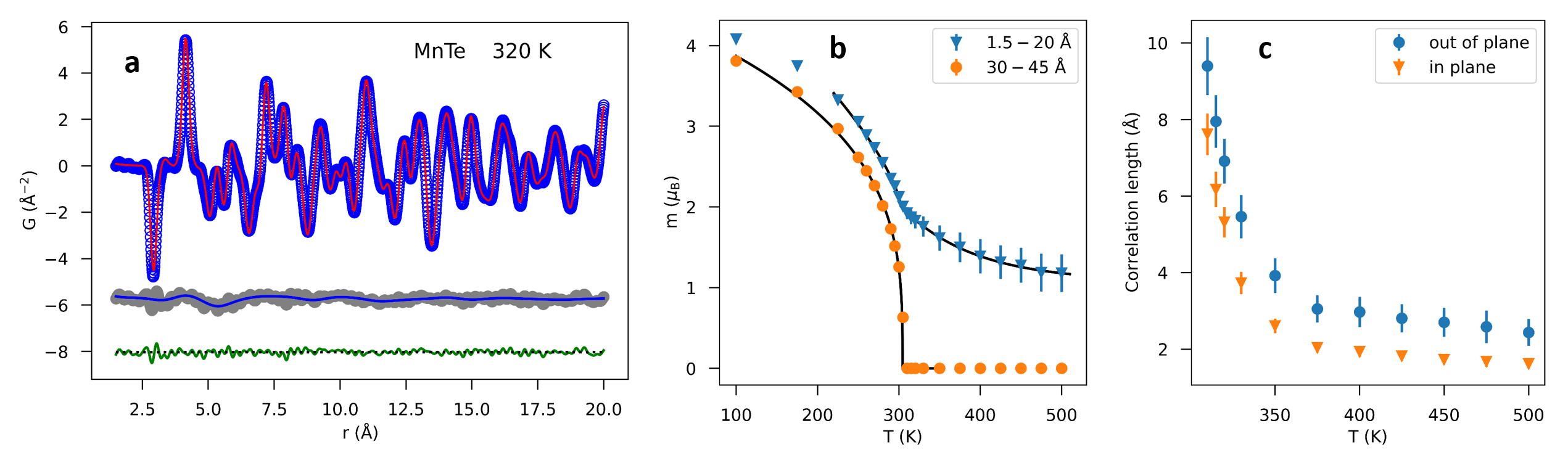}
	\caption{\label{fig:highTmPDFs} (a) Combined atomic and magnetic PDF fit for MnTe at 320~K, with the short-range AF correlations evidenced in the mPDF data (gray curve). (b) The ordered magnetic moment of MnTe as a function of temperature as determined from mPDF fits over a short fitting range (1.5-20~\AA, blue triangles, representing the ordered moment for nearest-neighbor spin pairs) and a longer fitting range (30-45~\AA, orange circles, representing the true long-range ordered moment). The solid curves are power law fits described in the main text. Representative fits over the longer fitting range are shown in the SI. (c) Temperature dependence of the best-fit correlation length along \textit{c} (blue circles) and within the \textit{ab} plane (orange triangles) in the paramagnetic regime, obtained from fits over 1.5 - 20~\AA. 
	}		
\end{figure*}

We performed similar fits for the data sets collected in the paramagnetic state. In Fig.~\ref{fig:highTmPDFs}(a), we display the atomic and magnetic PDF fits for MnTe at 320~K.
%%%%%%%%%%%%%%%%%
% Begin Figure
%%%%%%%%%%%%%%%%%
%%%%%%%%%%%%
% End Figure
%%%%%%%%%%%%
As seen in the lower set of curves, the mPDF component is reduced in magnitude and spatial extent compared to 100~K; nevertheless, it exhibits clear AF features in the low-\textit{r} region. These short-range AF correlations can be modeled adequately by including an isotropic exponential envelope of the form $\exp(-r/\xi)$ in the calculated mPDF, where the correlation length $\xi$ is a parameter in the fit. Qualitatively similar short-range features in the mPDF data are observed up to 500~K, the highest measured temperature. The locally ordered magnetic moment extracted from fits conducted over the ranges 1.5 – 20~\AA\ and 30 – 45~\AA\ (see SI) are shown in Fig.~\ref{fig:highTmPDFs}(b) as the blue triangles and orange circles, respectively. The latter fitting range reveals an abrupt onset of long-range AF correlations at 307~K that can be well described as a continuous phase transition with a critical exponent of $\beta=0.30 \pm 0.03$. In contrast, the short-range fits indicate that a significant fraction of the bare magnetic moment remains correlated among near neighbors up to at least 500~K. The thermal evolution of the locally ordered moment is reminiscent of a continuous phase transition biased by an applied field~\cite{blund;b;micm01}, suggesting that the local field created by neighboring spins plays the role of an applied field. Above \TN, the locally ordered moment can be well described by exponential decay that asymptotically approaches a constant value of 1.1~\muB\ with a characteristic temperature scale of 90~K, as shown by the solid curve. Additional details about the mPDF fits are given in the SI.

In light of the anisotropic correlation length observed in the \deltampdf\ data, we performed an additional set of mPDF fits above \TN\ over the range 1.5 - 20~\AA\ in which we implemented a model with distinct correlation lengths along \textit{c} (\xic)  and in the \textit{ab} plane (\xiab). At all temperatures above \TN, the best-fit values of \xic\ were longer than \xiab\ by approximately 50\% [see Fig.~\ref{fig:highTmPDFs}(c)], confirming the \deltampdf\ results and demonstrating that even a powder sample is sufficient for detecting spatially anisotropic correlations. As seen in Fig.~\ref{fig:highTmPDFs}, both correlation lengths show a rapid decrease with temperature from \TN\ to about 350~K, but remain fairly constant at higher temperatures, similar to the local magnetic moment in Fig.~\ref{fig:highTmPDFs}(b).~\footnote{We note that the magnetic correlation lengths above 350~K determined from the 1D mPDF fits are on the order of a few Angstroms, which is smaller than the value obtained by Zheng et al~\cite{zheng;sadv19} (~ 2.5 nm). This disparity is partially ameliorated by multiplying the mPDF result by $\pi$ to account for the different definitions of the magnetic correlation length used in the two works (see for example the Supplemental Material of Z. Dun \textit{et al}, \textit{Phys. Rev. B} \textbf{103}, 064424 (2021)). The remaining discrepancy is attributed to the fact that the 1D mPDF data effectively integrates over all neutron energy transfers, yielding the instantaneous spin-spin correlation function, while the previous work selected a narrow energy band around the elastic channel for calculating the correlation length, yielding the time-averaged correlation length. This is consistent with the longer correlation lengths obtained from the \deltampdf\ analysis when the statistical chopper was applied to obtain just the elastic scattering compared to the result using the total scattering.}

\subsection{\textit{Ab initio} theory: Spin correlation functions in the paramagnetic regime}	
\begin{figure}
	
	\includegraphics[width=75mm]{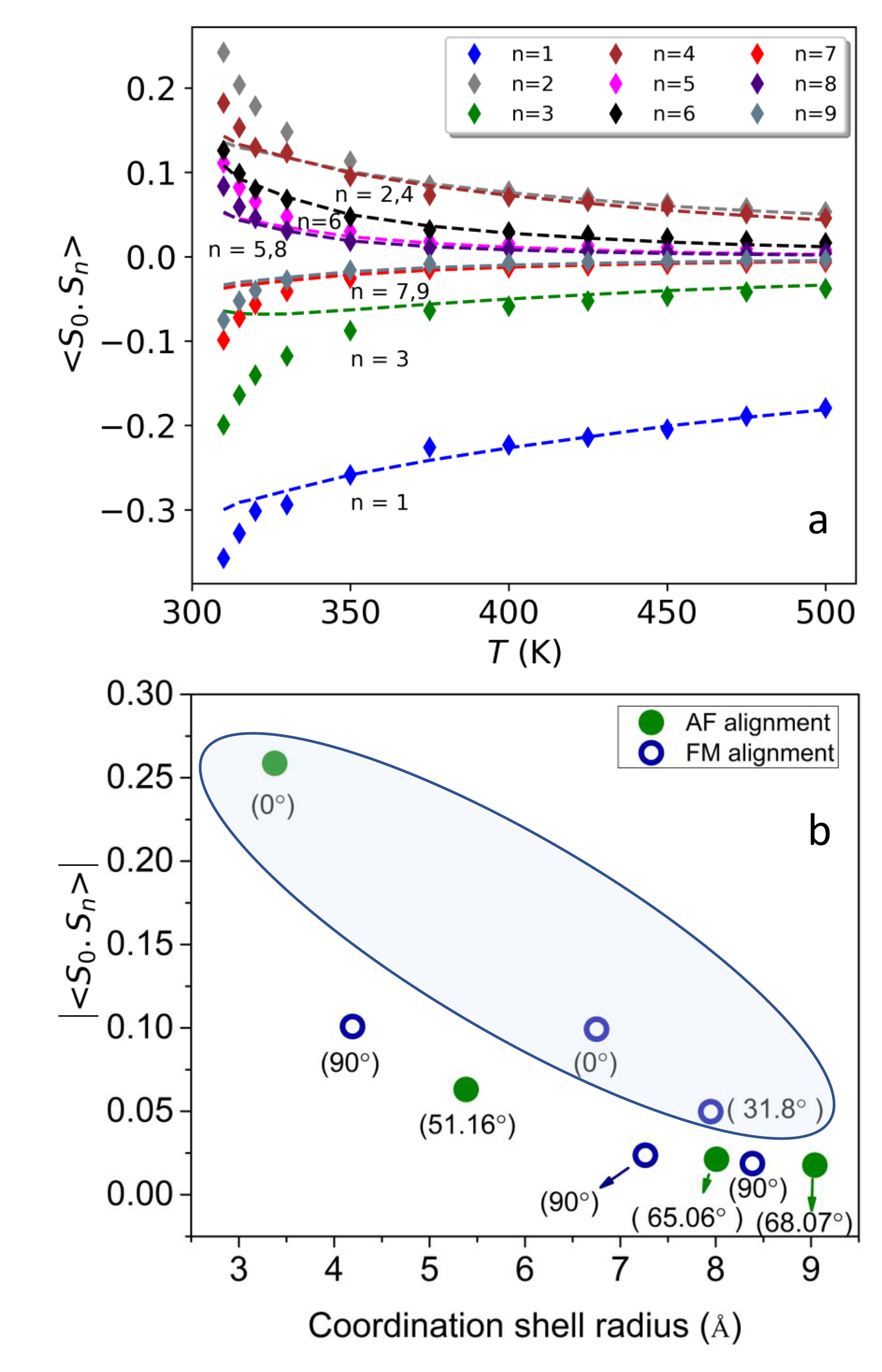}
	\caption{\label{fig:spinCorrels} (a) Theoretical and experimental spin correlation functions $\langle \mathbf{S}_0 \cdot \mathbf{S}_n \rangle$ versus temperature for the first nine nearest neighbors. The dotted line represents the theoretical and symbols represent the experimental magnetic correlations. (b) $\langle \mathbf{S}_0 \cdot \mathbf{S}_n \rangle$ versus coordination shell radius for the calculations at 350 K. Filled (open) symbols represent antiferromagnetic (ferromagnetic) alignment. The angle shown for each data point is the angle to the \textit{c} axis formed by each spin pair comprising the respective coordination shells. The shaded oval highlights anomalously strong correlations (see main text).
	}		
\end{figure}

The observed short-range magnetic correlations invite comparison with theoretical predictions. The spin correlation functions for the first nine coordination shells were calculated \textit{ab initio} as a function of temperature in the paramagnetic regime using the DLM-DFT-SIC approach~\cite{hughe;njp08}. The dashed curves in Fig.~\ref{fig:spinCorrels}(a) show the correlation functions $\langle \mathbf{S}_0 \cdot \mathbf{S}_n \rangle$, where $\mathbf{S}_0$ is an arbitrary spin, $\mathbf{S}_n$ is a spin in the $n^{\mathrm{th}}$ coordination shell relative to $\mathbf{S}_0$, and the angled brackets represent an average over all $\mathbf{S}_n$ in the shell.
%%%%%%%%%%%%%%%%%
% Begin Figure
%%%%%%%%%%%%%%%%%
%%%%%%%%%%%%
% End Figure
%%%%%%%%%%%%
The spins in the calculation have unit length. Negative and positive values correspond to net antiferromagnetic and ferromagnetic correlations, respectively. Corresponding spin correlation functions extracted from the 1D mPDF fits are displayed as diamond symbols in Fig.~\ref{fig:spinCorrels}(a). The experimental and theoretical results show remarkable agreement above 350~K, demonstrating that the DLM-DFT-SIC approach can be used to predict high-temperature magnetic correlations in MnTe with quantitative accuracy. The agreement is worse at lower temperature, where critical behavior associated with the phase transition becomes more dominant. The anisotropic nature of the spin correlations, which is captured naturally by the \textit{ab initio} calculations, is illustrated in Fig.~\ref{fig:spinCorrels}(b). The absolute value of each $\langle \mathbf{S}_0 \cdot \mathbf{S}_n \rangle$ calculated at 350~K is plotted as a function of coordination shell radius. If the spin correlations were to decay isotropically with distance, then the calculated correlation functions would lie along a smooth, monotonic curve that decreases with shell radius. However, the results are non-monotonic with distance, and we instead observe two distinct bands of correlation functions: a band of comparatively stronger correlations (indicated by the shaded oval) comprised of coordination shells where the spin separation vectors lie along the \textit{c} direction or close to it, and a band of weaker correlations with spin pairs located more in the $ab$ plane.

\section{Discussion}
The results presented here provide a unique, real-space picture of the short-range magnetic correlations in the correlated paramagnet state of MnTe. The \deltampdf\ and 1D-mPDF analyses reveal nanometer-scale, spatially anisotropic correlations that persist up to at least 500~K. The relative stability of the correlation length and the locally ordered moment above 350~K suggests that these short-range spin correlations remain present to much higher temperatures than measured in the current work, supporting the relevance of the paramagnon drag scenario at elevated temperatures where $zT$ approaches 1 in doped samples.

The newly revealed anisotropy of the correlations, i.e. the fact that $\xi_{c} > \xi_{ab}$, can be viewed as a natural consequence of the relative strength of the AF out-of-plane exchange interaction $J_1$ compared to the weaker ferromagnetic in-plane interaction $J_2$, both of which have been determined from inelastic neutron scattering data~\cite{szusz;prb06} and calculations~\cite{mu;prm19}. The discrepancy between $J_1$ and $J_2$ can be rationalized from a structural viewpoint in the context of superexchange theory~\cite{ander;63} by noting that along $c$, the Mn-Mn distance is $\sim$3.36~\AA\ and the Mn-Te-Mn exchange pathway forms an angle of $\sim$70$^{\circ}$ (favoring strong antiferromagnetism), while within the $ab$ plane, the Mn-Mn distance is $\sim$4.15~\AA\ and the Mn-Te-Mn angle is $\sim$90$^{\circ}$ (favoring weaker ferromagnetism). Thus, MnTe possesses a magnetic character somewhat similar to that of strongly interacting AF chains. The presence of anisotropic exchange interactions was recently proposed as a favorable ingredient for paramagnon-enhanced thermopower~\cite{polas;crps21}, as supported by the present work on MnTe.

The excellent agreement between the observed short-range spin correlations and the calculated correlations from the \textit{ab initio} DLM-DFT-SIC approach is notable for two reasons. First, it demonstrates that we now have a quantitatively accurate, first-principles model of the magnetic properties of MnTe in the high-temperature regime, where thermoelectric applications would be most likely. This lays the groundwork for future theoretical investigations of MnTe, including potential ways to improve the thermoelectric performance by paramagnon engineering. Second, these results further establish the DLM-DFT-SIC approach as a versatile and reliable method of predicting finite-temperature magnetic properties, having now seen success in numerous materials with diverse magnetic and electronic properties~\cite{staun;prl04,hughe;n07,ostan;prl07,aldou;prb12, staun;prb14,frand;prl16,patri;prm19}. Specific to the search for high-performance thermoelectric systems, this \textit{ab initio} framework can be used to screen candidate materials for the possibility of short-range magnetic correlations that may enhance $zT$ through paramagnon drag.

\section{Conclusion}
In summary, we have performed mPDF analysis and \textit{ab initio} calculations of MnTe to reveal the real-space nature of the short-range magnetic correlations responsible for enhancing $zT$ above \TN. We have shown that robust, short-range magnetic correlations persist on the nanometer scale well above \TN, consistent with the paramagnon picture proposed to explain the high value of $zT$. Further, we have demonstrated that these magnetic correlations are anisotropic, with a longer correlation length out of the plane than in plane, a consequence of anisotropic exchange interactions. \textit{Ab initio} calculations using the DLM-DFT-SIC method predict the observed anisotropic magnetic correlations with quantitative accuracy, validating this approach for the theoretical study of magnetically enhanced thermoelectric candidates. By offering unprecedented insight into the nature of the short-range magnetic correlations responsible for the enhanced thermoelectric response in MnTe, these results provide a vital benchmark for future efforts to quantify, understand, and optimize spin-driven thermoelectric enhancement due to short-range magnetic order in MnTe and other magnetic semiconductors.

\section{Experimental Methods}		
Polycrystalline samples of MnTe and Mn$_{0.98}$Na$_{0.02}$Te were prepared by mixing Mn powder (99.95\%), Te pieces (99.999\%), and Na pieces (99.99\%) in an argon glove box, sealing the mixture in an evacuated quartz ampoule, and holding it at 950$^{\circ}$C for 6 hours. The ampoule was then quenched in cold water and further annealed at 650$^{\circ}$C for 72 hours, following Ref.~\onlinecite{xu;jmchema17}. The samples were ground into a fine powder in a mortar and pestle inside the glove box. Temperature-dependent neutron total scattering experiments were performed on pure MnTe as well as Na-doped MnTe at the Nano Scale-Ordered Material Diffractometer (NOMAD) at the Spallation Neutron Source (SNS) of Oak Ridge National Laboratory (ORNL)~\cite{neuef;nimb12}. The powder samples were loaded in quartz capillaries inside an argon glove box. The total scattering patterns were collected at several temperatures between 100 K to 500 K. The total neutron scattering data were reduced and transformed with $Q_{\mathrm{max}}= 25$~\AA$^{-1}$ using the automatic data reduction scripts at NOMAD. The PDF analysis was performed using PDFgui~\cite{farro;jpcm07} and the magnetic PDF patterns were analyzed using the diffpy.mpdf package in the DIFFPY suite~\cite{juhas;aca15}. Magnetic measurements of powder samples were collected using vibrating sample magnetometer (VSM) instrument from Quantum Design. Both zero field cool (ZFC) and field cool (FC) measurements as a function of temperature were collected in the temperature range 100 – 400 K. Basic magnetic and structural characterization data are shown in the SI.

A single crystal of MnTe was grown out of Te flux by keeping a mixture of Mn:Te=36:64 at 890$^{\circ}$C for 12 days. Single crystal neutron diffraction patterns were collected at various temperatures on the CORELLI instrument~\cite{ye;jac18} at the SNS using neutrons with incident energies between 10 and 200~meV. The crystal was oriented such that the $H0L$ reciprocal space plane coincided with the horizontal scattering plane. At each temperature, the sample was rotated 360$^{\circ}$ around the vertical axis in steps of 3$^{\circ}$, with data collected for 3 minutes at each angle. The data were merged together and symmetrized using Mantid~\cite{arnol;nima14}. The diffuse magnetic scattering above \TN\ was isolated by subtracting the diffraction data collected around 300~K from the higher-temperature data sets, with the KAREN algorithm~\cite{weng;jac20} as implemented in Mantid used to fill in sharp negative holes in the data resulting from the subtraction of the magnetic Bragg peaks. The \deltampdf\ patterns were generated via three-dimensional Fourier transformation of the scattering data in Mantid using the DeltaPDF3D module~\cite{weber;zk12}.

\textbf{Acknowledgements}
	
We thank Michelle Everett, Jue Liu, and J\"org Neuefeind for their support at the NOMAD beamline and James Torres for assistance during the CORELLI beamtime at ORNL. Work by R.B., P.H., and B.A.F. was supported by the U.S. Department of Energy, Office of Science, Office of Basic Energy Science through Award No. DE-SC0021134. J.C. acknowledges support from the College of Physical and Mathematical Sciences at Brigham Young University. Work by J.B.S. was supported by the UK Engineering and Physical Sciences Research Council, Grant No. EP/M028941/1. Work at ORNL by JY and MAM (synthesis and magnetometry) and by MEM and RPH (neutron scattering) was supported by the U.S. Department of Energy, Office of Science, Office of Basic Energy Science, Materials Science and Engineering Division. This study used resources at the Spallation Neutron Source (SNS), a DOE Office of Science User Facility operated by the Oak Ridge National Laboratory.

\end{document}

\title{
	Supplementary Information: Real-space visualization of short-range antiferromagnetic correlations in a magnetically enhanced thermoelectric
}

	\author{Raju Baral}
	\affiliation{ %
		Department of Physics and Astronomy, Brigham Young University, Provo, Utah 84602, USA.
	} %
	
	\author{Jacob Christensen}
	\affiliation{ %
		Department of Physics and Astronomy, Brigham Young University, Provo, Utah 84602, USA.
	} %
	
	\author{Parker Hamilton}
	\affiliation{ %
		Department of Physics and Astronomy, Brigham Young University, Provo, Utah 84602, USA.
	} %

	\author{Feng Ye}
	\affiliation{ %
		Neutron Scattering Division, Oak Ridge National Laboratory, Oak Ridge, Tennessee 37831, USA.
	} %
	
	\author{Karine Chesnel}
	\affiliation{ %
		Department of Physics and Astronomy, Brigham Young University, Provo, Utah 84602, USA.
	} %

	\author{Taylor D. Sparks}
	\affiliation{ %
		Department of Materials Science and Engineering, University of Utah, Salt Lake City, Utah 84112, USA.
	} %

	\author{Rosa Ward}
	\affiliation{ %
		Department of Materials Science and Engineering, University of Utah, Salt Lake City, Utah 84112, USA.
	} %

	\author{Jiaqiang Yan}
	\affiliation{ %
		Materials Science \& Technology Division, Oak Ridge National Laboratory, Oak Ridge, Tennessee 37831, USA.
	} %

	\author{Michael A. McGuire}
	\affiliation{ %
		Materials Science \& Technology Division, Oak Ridge National Laboratory, Oak Ridge, Tennessee 37831, USA.
	} %
	
	\author{Michael E. Manley}
	\affiliation{ %
		Materials Science \& Technology Division, Oak Ridge National Laboratory, Oak Ridge, Tennessee 37831, USA.
	} %
	
	\author{Julie B. Staunton}
	\affiliation{ %
	Department of Physics, University of Warwick, Coventry CV4 7AL, United Kingdom.
	}%
	
	\author{Rapha\"el P. Hermann}
	\email{hermannrp@ornl.gov}
	\affiliation{ %
		Materials Science \& Technology Division, Oak Ridge National Laboratory, Oak Ridge, Tennessee 37831, USA.
	} %

	\author{Benjamin A. Frandsen}
	\email{benfrandsen@byu.edu}
	\affiliation{ %
		Department of Physics and Astronomy, Brigham Young University, Provo, Utah 84602, USA.
	} %

\maketitle

\section{Basic magnetic and structural characterization}
\begin{figure}
	\includegraphics[width=160mm]{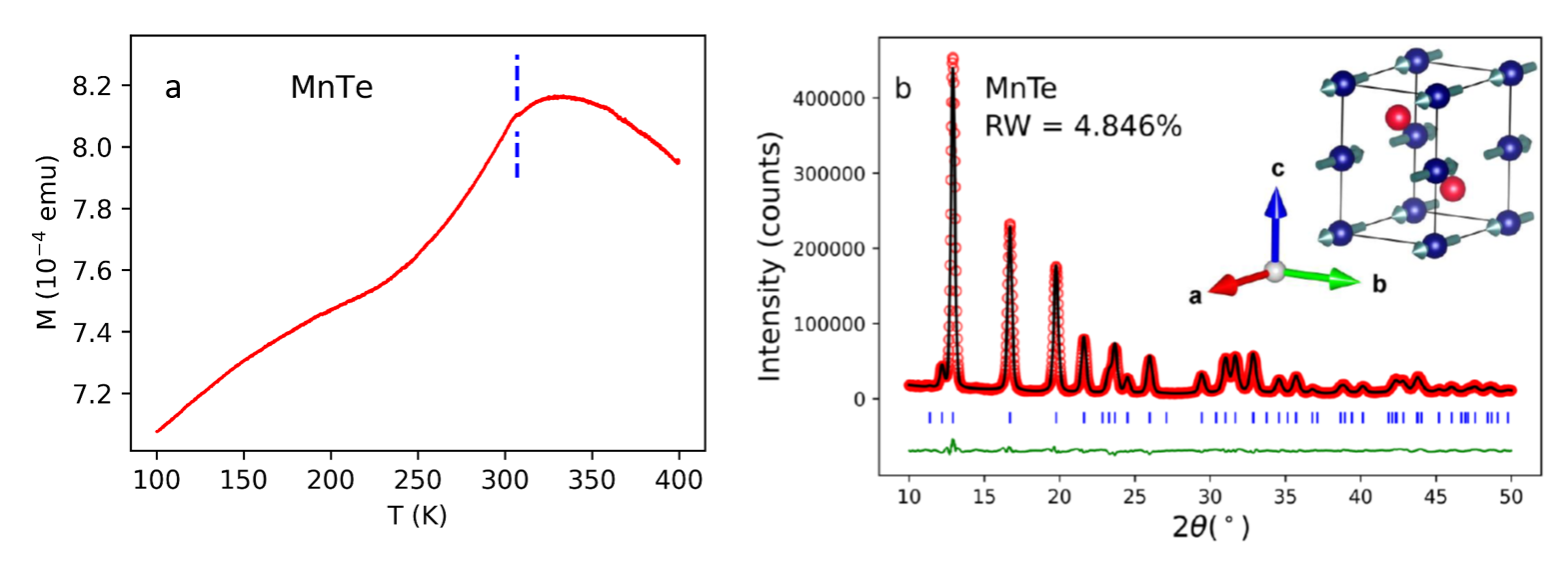}
	\caption{\label{fig:basic} (a) Field-cooled measurement of MnTe performed under a field of $H = 1000$~Oe, where the magnetization is plotted as a function of temperature while warming up. The dashed vertical line marks $T_{\mathrm{N}}=307$~K. (b) The Rietveld refinement pattern (black curve) plotted on the top of X-ray diffraction (red symbols) of MnTe. The green curve shows the fit residual and the blue vertical ticks are the Bragg positions. Inset: Crystal and magnetic structure of MnTe. Blue and red spheres represent Mn and Te atoms, respectively.}
		
\end{figure}

In Fig.~\ref{fig:basic}(a), we display the field-cooled (FC) measurement of MnTe performed under a field of $H=1000$~Oe, showing the magnetization $M(T)$ as a function of temperature while warming up. The visible kink in the curve just above 300~K confirms the expected magnetic phase transition. In Fig.~\ref{fig:basic}(b), we show a Rietveld refinement of the x-ray diffraction (XRD) data collected on a laboratory XRD instrument using the published crystal structure of MnTe. The Rietveld refinement of powder x-ray diffraction was carried out using GSAS-II~\cite{toby;jac13}.The results confirm the phase purity of the sample.
%%%%%%%%%%%%%%%%%
% Begin Figure
%%%%%%%%%%%%%%%%%
%%%%%%%%%%%%
% End Figure
%%%%%%%%%%%%

\section{Neutron total scattering structure function}
\begin{figure}
	\includegraphics[trim=35mm 55mm 40mm 15mm, clip,width=180mm]{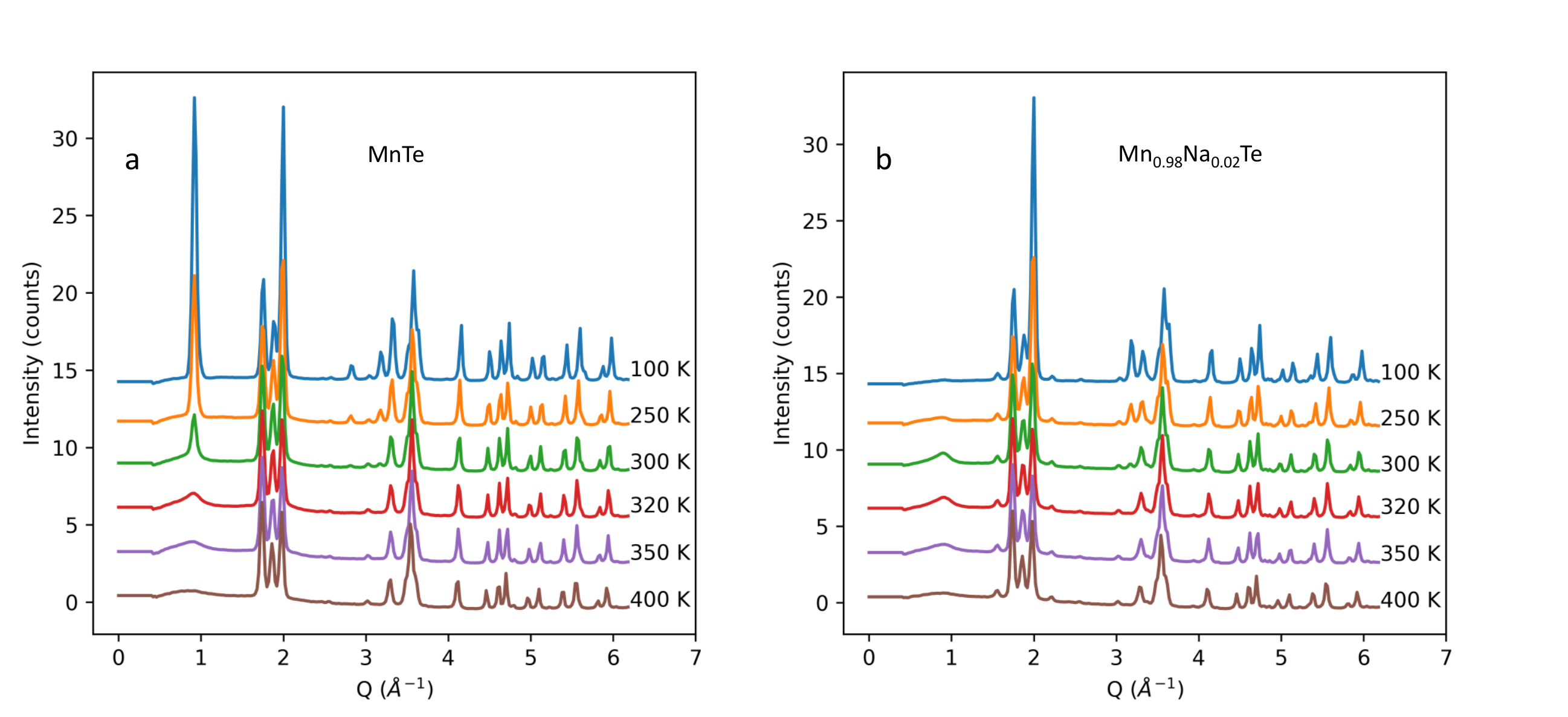}
	\caption{\label{fig:SofQ} (a,b) $S(Q)$ curve for various temperatures for MnTe and Mn$_{0.98}$Na$_{0.02}$Te respectively.}
		
\end{figure}

The neutron total scattering structure function $S(Q)$ is shown for MnTe and Na-MnTe at various temperatures in Fig.~\ref{fig:SofQ}. The $S(Q)$ plots of pure MnTe in panel (a) show prominent magnetic Bragg peaks below \TN, e.g. around 0.9~\AA$^{-1}$ and 1.9~\AA$^{-1}$. As the temperature increases above \TN, these sharp peaks become broad and diffuse features due to short-range magnetic correlations. Fig.~\ref{fig:SofQ}(b) shows equivalent $S(Q)$ plots for Mn$_{0.98}$Na$_{0.02}$Te. The most notable difference is the absence of the strong magnetic peak around 0.9~\AA$^{-1}$ at low temperature, which can be attributed to the reorientation of the spins from an in-plane direction in pure MnTe to along the $c$ axis in the doped compound. However, as the temperature is raised, some magnetic intensity is observed around 0.9~\AA$^{-1}$, indicating that the spin direction gains some component perpendicular to $c$, as also observed for Li doping~\cite{mosel;prm21}. Above \TN, similar diffuse scattering as in pure MnTe is observed.
%%%%%%%%%%%%%%%%%
% Begin Figure
%%%%%%%%%%%%%%%%%
%%%%%%%%%%%%
% End Figure
%%%%%%%%%%%%

\section{1D atomic and magnetic PDF fits}
One dimensional atomic and magnetic fits to the neutron PDF data were done using the following iterative fitting approach. We first performed an atomic PDF fit, using the published structure as a starting point for the refinement. The calculated atomic PDF corresponding to this fit was then subtracted from total PDF data, yielding a fit residual that was dominated by the experimental magnetic PDF signal but also contained any imperfections in the atomic PDF fit. This fit residual served as the input data for an initial mPDF refinement. A second atomic PDF fit was then conducted, but this time the input data was the total PDF data minus the best-fit calculated mPDF. The parameter values from the first atomic PDF fit were used as starting points for this next iteration. The reason for this second fit was to allow a more precise refinement of the structure against data from which the magnetic component had been largely removed. We then subtracted the best-fit atomic PDF from the original total PDF data and used the result as input data for a second mPDF fit. The calculated mPDF was then subtracted from the original data to perform a third atomic PDF fit, and finally a third mPDF fit was conducted in the same fashion. We found that the goodness-of-fit metric $R_w$ improved with each successive iteration, but additional fits beyond these three iterations resulted in no meaningful improvement. We also attempted a true corefinement of the atomic and magnetic structures together, but we found that the fits tended to terminate in local minima in this case, whereas the iterative approach with alternating atomic and magnetic PDF fits was more robust.

The atomic PDF fit parameters included the lattice parameters $a$ and $c$, the hexagonal symmetry-allowed anisotropic displacement parameters (ADPs), the $z$ coordinates of the Te and Mn atomic sites (also allowed to vary within the symmetry of the space group), the linear correlated motion correction parameter, and an overall scale factor. The instrumental parameters $Q_{damp}$ and $Q_{broad}$ were fixed to values determined from calibration fits to a silicon standard.

The magnetic model assumed the published antiferromagnetic structure but included as free parameters the paramagnetic scale factor (corresponding to the self-scattering component of the magnetic differential scattering cross section~\cite{frand;aca15}), the ordered scale factor (related to the square of the locally ordered magnetic moment), and the direction of the magnetic moments as defined by the angle $\theta$ from the $c$ axis. We note that the paramagnetic scale factor is based on the isotropic magnetic form factor for Mn$^{2+}$ tabulated in Supplementary Ref.~\onlinecite{wilso;b;itc95}. Since the self-scattering contribution is restricted largely to $r<1$~\AA~\cite{frand;aca15}, which was excluded from our fits, this parameter has only a minor effect on the fit, but was nevertheless included for completeness. For the case of fits corresponding to short-range magnetic correlations (as opposed to long-range magnetic order), we additionally included two parameters quantifying the exponential correlation length along the $c$ axis and in the $ab$ plane called $\xi_c$ and $\xi_{ab}$, respectively. For a given pair of magnetic moments with a separation vector $\mathbf{r}_{\mathrm{sep}}=r_{\mathrm{sep}}\hat{n}$, where $r_{\mathrm{sep}}$ is the distance separating them and $\hat{n}$ is the direction of the separation vector, the correlation length was calculated as
\begin{align}
    \xi(\hat{n})=\left( \frac{n_x^2+n_y^2}{\xi_{ab}^2}+\frac{n_z^2}{\xi_c^2} \right)^{-\frac{1}{2}},
\end{align}
where $n_x,n_y,n_z$ are the components of $\hat{n}$ in Cartesian coordinates. Thus, the correlation length corresponds to the surface of an ellipsoid in real space with semi-axes of $\xi_{ab}$ along the Cartesian axes $x$ and $y$ and $\xi_c$ along the Cartesian $z$ direction. Relative to whichever moment was arbitrarily taken to be at the origin, then, the second magnetic moment in the pair has a magnitude of $m_0 \exp [-r_{\mathrm{sep}}/\xi(\hat{n})]$, where $m_0$ is the magnitude of the moment at the origin.

\begin{figure*}
	\includegraphics[width=140mm]{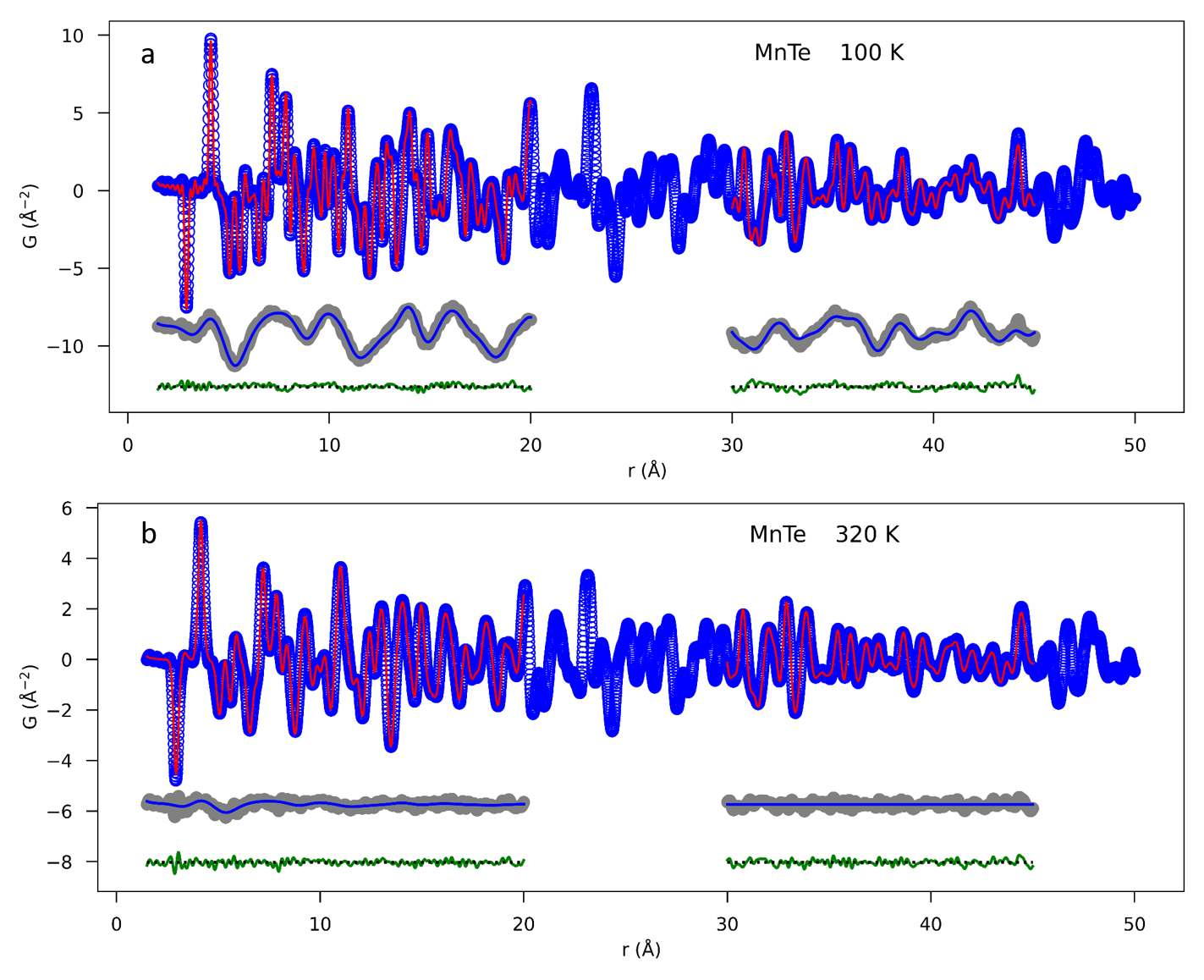}
	\caption{\label{fig:mPDFtworanges} Combined atomic and magnetic PDF fits at 100~K (a) and 320~K (b) for the two different fitting ranges discussed in the main text, 1.5 -- 20~\AA\ and 30 -- 45~\AA.}
\end{figure*}

In Fig.~\ref{fig:mPDFtworanges}, we display the combined atomic and magnetic PDF fits at 100~K and 320~K for the two fitting ranges discussed in the main text.
%%%%%%%%%%%%%%%%%
% Begin Figure
%%%%%%%%%%%%%%%%%
%%%%%%%%%%%%
% End Figure
%%%%%%%%%%%%
The shorter fitting range captures the short-range magnetic correlations present above \TN, whereas the longer fitting range is representative of the genuine long-range magnetic order.

We note that the default calibration at NOMAD has limited accuracy for $Q \lesssim 2.4$~\AA$^{-1}$. The $(0, 0, 1)$ magnetic Bragg peak is located at $Q = 2\pi/c \approx 0.93$~\AA$^{-1}$, meaning its apparent $Q$ position in the data is affected by the suboptimal calibration. This introduces artifacts into the $S(Q)$ and $G(r)$ patterns. To correct for this, we recalibrated the $S(Q)$ data in the range $0.5 \le Q \le 2.4$~\AA$^{-1}$ using the known position of the magnetic Bragg peak at low temperature, which improved the quality of the data and allowed more quantitatively accurate fits.

\section{Spin direction determined from mPDF fits}
\begin{figure}
	\includegraphics[width=70mm]{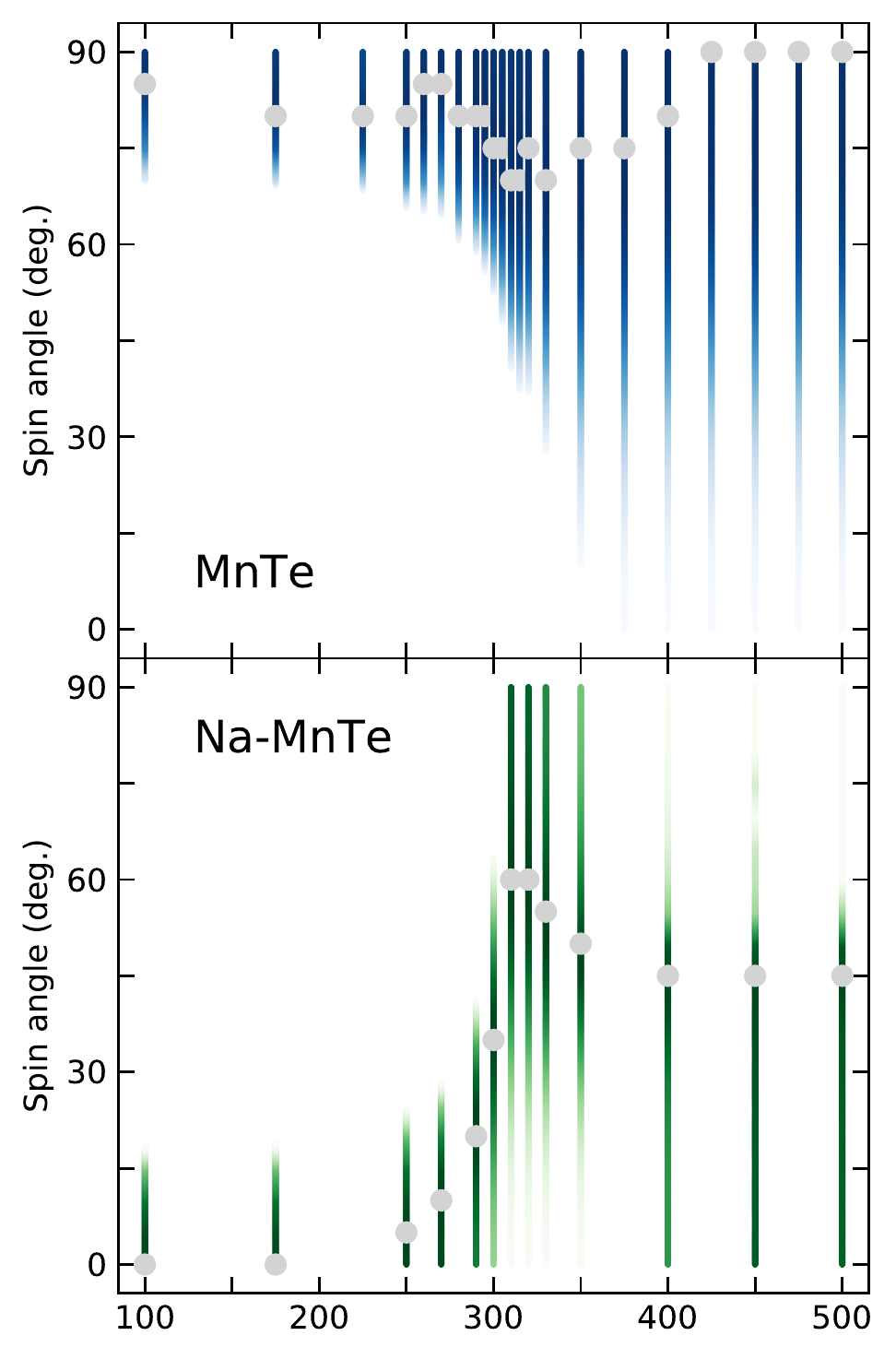}
	\caption{\label{fig:spinangles} Best-fit spin directions for MnTe (top) and Na-doped MnTe (bottom). The vertical axis on the figure indicates the angle between the $c$ axis of the unit cell and the direction of the ordered magnetic moment. The gray circles indicate the angle yielding the best fit for each temperature. The shaded bars represent a type of confidence interval as explained in the main body of the text. }
		
\end{figure}

To determine the best-fit spin direction from the mPDF data at each temperature, we performed a series of mPDF fits where the angle $\theta$ between the spin and the $c$ axis was set to values ranging from 0$^{\circ}$ to 90$^{\circ}$ in 5$^{\circ}$ steps. The azimuthal angle within the $ab$ plane makes no difference for the orientationally averaged 1D mPDF due to the hexagonal symmetry, so it was fixed to zero (corresponding to the [100] direction) for simplicity. The goodness-of-fit metric \Rw\ was then extracted from each fit, and the angle corresponding to the lowest value of \Rw\ (hence the best overall fit) was designated as the best-fit angle. These are shown as the gray circles in Fig.~\ref{fig:spinangles}.
%%%%%%%%%%%%%%%%%
% Begin Figure
%%%%%%%%%%%%%%%%%
%%%%%%%%%%%%
% End Figure
%%%%%%%%%%%%
The shaded bars underneath the gray circles indicate the range of spin angles that result in \Rw\ values within 0.5\% of the optimal \Rw\ value, where 0.5\% was chosen as a reasonable threshold for meaningful differences between fits. The hue of the colored bar lightens proportionally with the increase in \Rw\ from the optimal value, fading to white at the 0.5\% threshold. Thus, these shaded bars represent a type of confidence interval for the best-fit spin direction as determined by the mPDF fits. Below about 250~K, the pure and doped compounds show a strong preference for an in-plane orientation ($\theta = 90^{\circ}$) and out-of-plane orientation ($\theta = 0^{\circ}$), respectively, with narrow confidence intervals. As the temperature increases further, the confidence intervals for both compounds widen due to the reduced weight of the magnetic component in the total PDF. Additionally, the best-fit spin angle for the doped compound cants away from the $c$ axis and assumes an angle of 60$^{\circ}$ at \TN, i.e. 30$^{\circ}$ above the $ab$ plane. Above \TN, the 0.5\% \Rw\ threshold covers the entire angular range, although there is still a slight improvement to the fit for spin angles closer to the $ab$ plane for the pure compound and closer to the $c$ axis for the doped compound. In any case, the strong directional preference of the ordered spins that is present deep in the ground state is largely absent from the high-temperature mPDF data presented here.

\section{Obtaining and modeling the \deltampdf}
\begin{figure}
	\includegraphics[width=100mm]{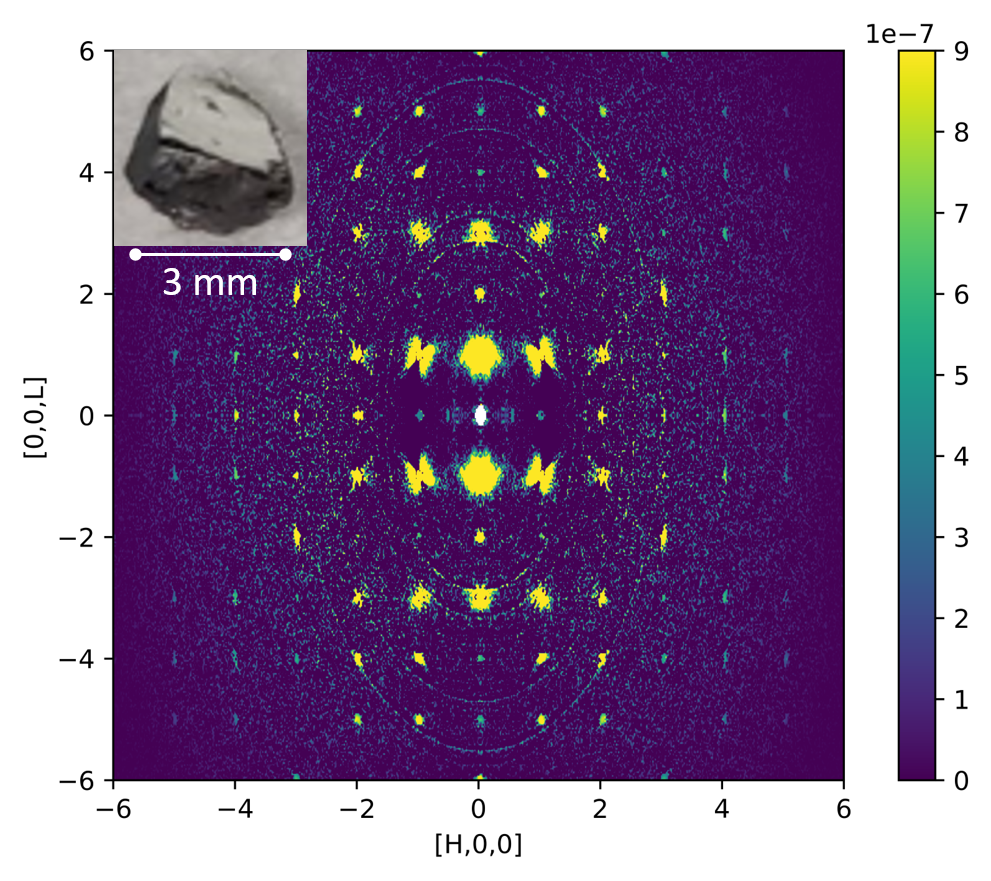}
	\caption{\label{fig:intermediate} Diffuse magnetic scattering in MnTe at 340~K. Inset: Photo of the single crystal used for the experiment.}
		
\end{figure}

Single crystal neutron diffraction experiments were carried out on the CORELLI instrument to measure the scattered intensity in large volumes of reciprocal space. The sample had a mass of 96.4~mg with all three dimensions similar in size ($\sim$3~mm), as seen in the inset of Fig.~\ref{fig:intermediate}. At each temperature, the sample was rotated through 360$^{\circ}$ in steps of 3$^{\circ}$. Due to the wavelength range of the incident neutrons ($\sim 0.64 - 2.86$~\AA), this angular step size allows continuous coverage of reciprocal space. The resulting 120 scattering patterns were then merged together, symmetrized according to the point group $6/mmm$, and normalized by the scattering from a vanadium reference sample, all within Mantid~\cite{arnol;nima14}.

To isolate the diffuse magnetic scattering above \TN, we subtracted the data collected at 300~K (in the ordered state) from the data collected at 340~K. This largely removed the nuclear Bragg peaks and background signal from the instrument and sample can, but it also created sharp negative features at the positions of the magnetic Bragg peaks. To eliminate these features and other artifacts of the temperature subtraction, we utilized the KAREN algorithm~\cite{weng;jac20} as implemented in Mantid, which identifies statistical outliers and replaces them with a smooth interpolating function. This resulted in the intermediate scattering pattern shown in Fig.~\ref{fig:intermediate}, where we clearly observe diffuse magnetic scattering centered around the original positions of the magnetic Bragg peaks present below \TN.
%%%%%%%%%%%%%%%%%
% Begin Figure
%%%%%%%%%%%%%%%%%
%%%%%%%%%%%%
% End Figure
%%%%%%%%%%%%

This intermediate scattering pattern was passed through the DeltaPDF3D module~\cite{weber;zk12} in Mantid, a three-dimensional fast Fourier transform (FFT) algorithm that yields the \deltampdf. To reduce noise in the FFT, we restricted the input scattering data to a sphere of reciprocal space with a radius of approximately 9~\AA$^{-1}$. This does not exclude any meaningful magnetic scattering because the magnetic form factor suppresses scattering at high $\mathbf{Q}$. 

\begin{figure}
	\includegraphics[width=150mm]{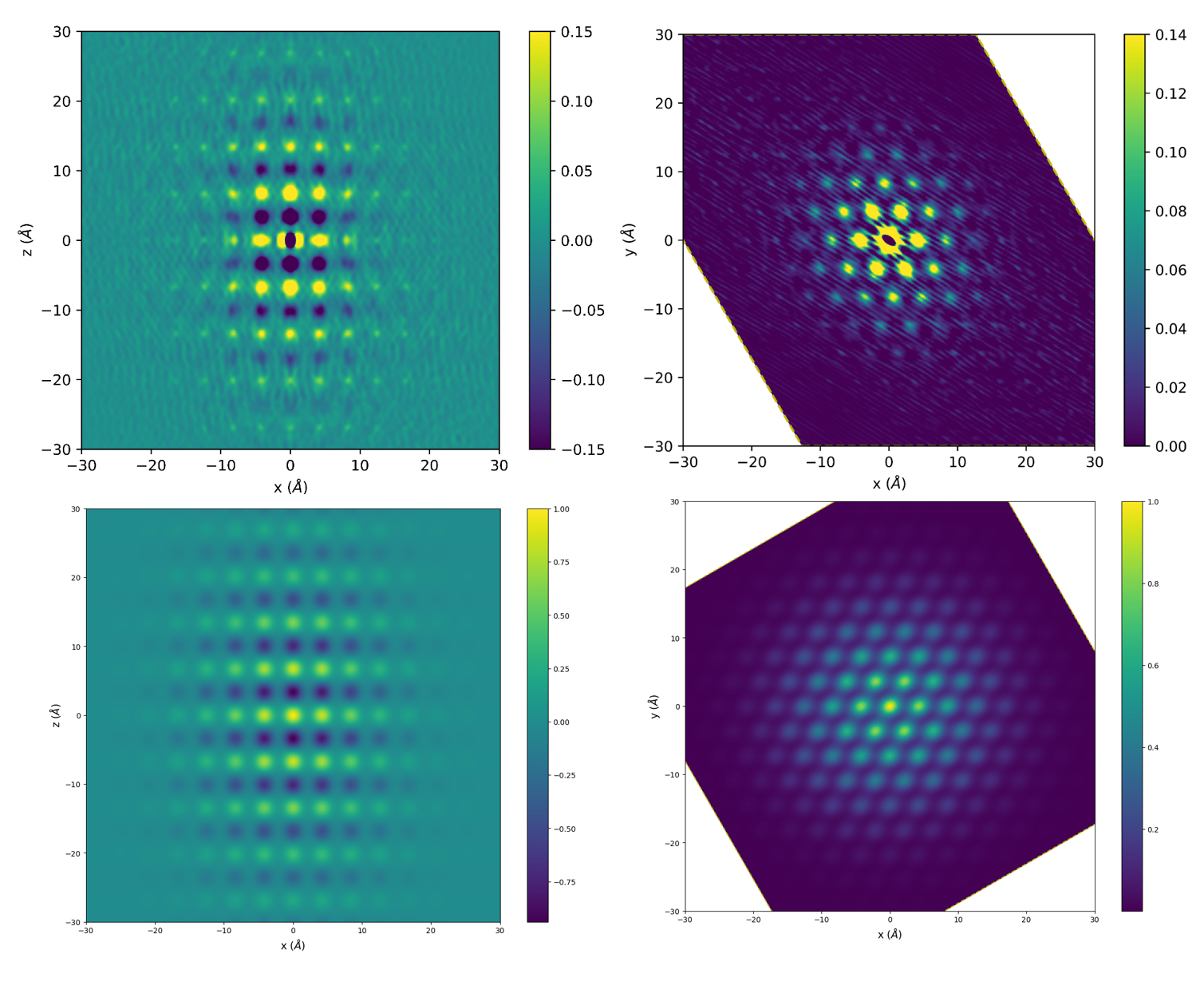}
	\caption{\label{fig:exp-calc} (a, b) Experimental \deltampdf\ pattern at 340~K in the $(x,0,z)$ plane (a) and $(x,y,0)$ plane (b). (c, d) Corresponding calculated \deltampdf\ patterns using the anisotropic correlation lengths determined from fits to line cuts through the \deltampdf\ data as explained in the main manuscript.}
		
\end{figure}

In Fig.~\ref{fig:exp-calc}, we show slices of the \deltampdf\ at 340~K for the $(x,0,z)$ plane (also shown in the main manuscript) and the $(x,y,0)$ plane in panels (a) and (b), respectively. The alternating spin orientation along the $z$ direction is clearly seen as the alternating bright and dark spots in panel (a), while the ferromagnetic alignment within the $xy$ plane is illustrated in panel (b).
%%%%%%%%%%%%%%%%%
% Begin Figure
%%%%%%%%%%%%%%%%%
%%%%%%%%%%%%
% End Figure
%%%%%%%%%%%%
We show the corresponding calculated \deltampdf\ patterns in panels (c) and (d), where the model assumed the standard MnTe antiferromagnetic structure but with anisotropic correlation lengths applied, as discussed in the previous section. The correlation lengths were determined from fits to line cuts through the \deltampdf\ data (see main manuscript). We performed the calculations using a home-built extension of the diffpy.mpdf python package following the definition of the \deltampdf\ given in Ref.~\onlinecite{roth;iucrj18}.

\section{DLM-SIC-DFT calculations}
The calculations of the spin correlations were carried out using the disordered local moment - density functional theory (DLM-DFT) of finite-temperature magnetism \cite{gyorf;jppfm85,staun;prb14} in which strong electronic correlations, evident in transition metal oxides \cite{hughe;njp08}, lanthanide metals \cite{hughe;n07,mendi;prl17} and rare earth- transition metal permanent magnets \cite{patri;prb18,patri;prm19}, for example, are treated with the self-interaction correction \cite{luder;prb05}. The DLM-DFT incorporates effects of thermally induced magnetic excitations and can thus describe quantitatively both rare earth (RE) and transition metal (TM) magnetism at finite temperatures. From its {\it ab initio} Gibbs free energy of a magnetic material \cite{mendi;prb19} complex magnetic interactions can be extracted for use in further modelling. The interactions can turn out to have both components linking pairs of sites, which indicate the most stable magnetic phases including non-collinear, frustrated or long period states, and components multi-site in nature depending on the state and extent of magnetic order. No prior assumption of any specific effective spin Hamiltonian is made, and quantitatively accurate magnetic phase diagrams and also caloric effects can be obtained from this approach e.g. \cite{mendi;prb19,staun;prb14,mendi;prl17}.

When applied to MnTe, the theory accounts for the insulating gap and its persistence into the paramagnetic phase, the magnetic ordering temperature $T_N$, and the onset of the AF state below $T_N$. Our first-principles DLM-DFT calculations find the ground-state Mn-ion configuration to be Mn$^{2+}$ with five localized $d$ states constituting a half-filled shell in line with Hund's first rule. For the comparison with the experimental PDF data of this paper, the key quantity from the theory is the lattice Fourier transform (LFT), $\mathcal{S}^{(2)}_{a,b} ({\bf q})$, of the direct correlation function for the local Mn moments in the paramagnetic state. The magnetic correlation function in real space, $C_{0,n} = \langle {\bf s}_0 \cdot {\bf s}_n \rangle$, is determined from this fundamental quantity. The subscripts $a$,$b$ denote the sites within the unit cell; for MnTe, these are the two sites in the hexagonal NiAs unit cell occupied by Mn atoms. Here, ${\bf s}_0$ represents a classical Heisenberg spin-like local moment arbitrarily chosen to be at the origin, ${\bf s}_n$ represents a spin in the $n$th coordination shell, and the angled brackets denote an average over all the spins in that shell. For MnTe, calculations of $\mathcal{S}^{(2)}_{a,b} ({\bf q})$ were performed for a grid of 56 wavevectors, {\bf q}, in the irreducible segment of the Brillouin zone. We found the  $C_{0,n}$'s using an Onsager cavity field calculation \cite{staun;prl92} by solving the coupled integral equations, $C^{-1}_{a,b}({\bf q}) = [\delta_{a,b}-\beta (\mathcal{S}^{(2)}_{a,b} ({\bf q}) - \Lambda_a \delta_{a,b})]$ and $\Lambda_a = \int \sum_b \mathcal{S}^{(2)}_{a,b} ({\bf q}) C_{b,a}({\bf q}) d {\bf q}$ for the LFT of the magnetic correlation function, $C_{a,b}({\bf q})$. This pair of equations ensures that the sum rule $\langle s^2_a \rangle = 1$ is met. Furthermore, the real-space direct correlation quantities describe the magnetic exchange interactions, the $J$ 's, between the Mn spins on different shells.